\PassOptionsToPackage{table}{xcolor}
\documentclass[sigconf=true, nonacm=true]{acmart}
\usepackage{booktabs}

\AtBeginDocument{%
  \providecommand\BibTeX{{%
    \normalfont B\kern-0.5em{\scshape i\kern-0.25em b}\kern-0.8em\TeX}}}

\copyrightyear{2021}
\acmYear{2021}
\setcopyright{acmlicensed}\acmConference[GECCO '21 Companion]{2021 Genetic and Evolutionary Computation Conference Companion}{July 10--14, 2021}{Lille, France}
\acmBooktitle{2021 Genetic and Evolutionary Computation Conference Companion (GECCO '21 Companion), July 10--14, 2021, Lille, France}
\acmPrice{15.00}
\acmDOI{10.1145/3449726.3463218}
\acmISBN{978-1-4503-8351-6/21/07}

\usepackage{dsfont}
\usepackage{soul,longtable,lscape,subfigure} 
\usepackage{xcolor,hyperref,amsmath, booktabs}

\newcommand{\E}{\ensuremath{\operatorname{E}}}

\acmConference[GECCO '21]{the Genetic and Evolutionary Computation Conference 2021}{July 10--14, 2021}{Lille, France}
\acmPrice{15.00}
\acmISBN{978-x-xxxx-xxxx-x/YY/MM}

\begin{document}

\title{Is there Anisotropy in Structural Bias?}

\author{Diederick Vermetten}
\orcid{1234-5678-9012} 
\affiliation{
  \institution{LIACS, Leiden University}
  \country{The Netherlands}}
\email{d.l.vermetten@liacs.leidenuniv.nl}

\author{Anna V. Kononova}
\orcid{0002-4138-7024}
\affiliation{
  \institution{LIACS, Leiden University}
  \country{The Netherlands}}
\email{a.kononova@liacs.leidenuniv.nl}

\author{Fabio Caraffini}
\authornote{Corresponding author}
\orcid{0001-9199-7368} 
\affiliation{
  \institution{Institute of Artificial Intelligence, \\De Montfort University}
  \country{Leicester, UK}}
\email{fabio.caraffini@dmu.ac.uk}

\author{Hao Wang}
\orcid{1234-5678-9012} 
\affiliation{
  \institution{LIACS, Leiden University}
  \country{The Netherlands}}
\email{h.wang@liacs.leidenuniv.nl}

\author{Thomas B{\"a}ck}
\orcid{0001-6768-1478}
\affiliation{
  \institution{LIACS, Leiden University}
  \country{The Netherlands}}
\email{t.h.w.baeck@liacs.leidenuniv.nl}

\renewcommand{\shortauthors}{Vermetten et al.}

\begin{abstract}
    Structural Bias (SB) is an important type of algorithmic deficiency within iterative optimisation heuristics. However, methods for detecting structural bias have not yet fully matured, and recent studies have uncovered many interesting questions. One of these is the question of how structural bias can be related to anisotropy. Intuitively, an algorithm that is not isotropic would be considered structurally biased. However, there have been cases where algorithms appear to only show SB in some dimensions. As such, we investigate whether these algorithms actually exhibit anisotropy, and how this impacts the detection of SB. We find that anisotropy is very rare, and even in cases where it is present, there are clear tests for SB which do not rely on any assumptions of isotropy, so we can safely expand the suite of SB tests to encompass these kinds of deficiencies not found by the original tests.
    
    We propose several additional testing procedures for SB detection and aim to motivate further research into the creation of a robust portfolio of tests. This is crucial since no single test will be able to work effectively with all types of SB we identify. 
    
 
\end{abstract}


\maketitle

\section{Introduction}
The modern world has become more computationally daring. More and more complex optimisation problems need to be solved to facilitate the ever-growing technological boom. For the majority of these problems, solving them exactly is no longer possible computationally due to their dimensionality, complexity or computability. Luckily, in most situations, it is also no longer necessary: good heuristic optimisation methods deliver sufficiently good solutions. However, it does not mean that such solutions are readily obtainable: good heuristics require extensive computing resources and (long) computation time, careful design and tuning. These latter aspects are largely based on the experience of computer scientists who are capable of offering an efficient solution for a given problem.

Apart from recently emerging Deep Learning approaches that have tremendously grown in popularity over the recent years, iterative heuristic approaches deliver excellent and explainable results. A large body of research accumulated in the fields of classical iterative optimisation~\cite{nelder1965simplex, powell1978fast} and computational intelligence~\cite{goldberg1988genetic,kennedy1995particle, hansen2001self_adaptation_es, price2006differential} potentially allows better guided choices free of various deficiencies. 
While these iterative optimisation heuristics are very effective, their design and configuration is a challenging problem in itself, with many different aspects to consider in order to create an effective algorithm. 

One key aspect is the sampling of solutions within an iterative heuristic optimisation algorithm, which is clearly driven by the landscape of objective function, or more precisely, the differences in the values of (or derived from) the objective function of candidate solutions. Loosely following such `survival of the fittest' logic, the algorithm is moving in function's domain (slowly) improving the values of objective function and, thus, heuristically solving the problem. 

However, it has been established \cite{Kononova2015,Caraffini2019} that this mechanism is not the only force driving the search - iterative application of individual operators that make up the algorithm can produce their own bias, the so-called structural bias (SB) of the algorithm. In computation, it manifests itself as a nonuniform preference of different parts of domain regardless of the objective function. The resulting movement of search then becomes a superposition of two `forces': landscape and structural bias. Typically, the former one largely overpowers the latter one. However, structural bias can potentially \textit{hinder} the search in case the optima are located in the less `favoured' part of the domain. In this sense, structural bias is an \textit{algorithmic deficiency}\footnote{Since optima locations are by definition not known beforehand, we cannot use structural bias to our advantage by biasing the search towards them. Moreover, at present, mechanisms of formation of SB are not fully understood.}.

While some algorithms, notably Evolution Strategies~\cite{hansen2001self_adaptation_es}, propose design principles for the algorithm development which are theoretically motivated to be unbiased in several areas~\cite{HansenA14}, with the most prominent ones being the rotation invariance (essentially isotropy in Euclidean spaces) and the stationarity of the search under random selection (equivalent to unbiasedness on $f_0$\footnote{See Section~\ref{sec:sb_detect}}), these kinds of algorithms are the exception rather than the rule. As such, detecting presence of SB for a certain algorithm is still a useful technique for better understanding the design choices for different algorithms.
It has been shown \cite{Kononova2015,vStein2021_emergence} that structural bias is not easy to be identified as it intricately depends on the choice of algorithm's framework, operators and parameters. The proposed identification procedure \cite{Kononova2015,Kononova2020PPSN} involves executing a series of algorithm runs on a special test function and examining the locations of final solutions. This special function is designed to decouple the aforementioned superposition of forces by deactivating one of them, the landscape force. The collected data is examined by means of visual or statistical tests. However, at the current stage, statistical tests, typically preferent over the visual ones, have been shown to have computational problems. One of such problems stems from their application on the per dimension basis to account for potential differences in how generic algorithms treat each dimension. \textit{What should be concluded if statistical test returns a different verdict for different dimensions?} \textit{Would it be reasonable to assume structural bias can indeed exhibit anisotropy in presence, strength or kind, i.e. different dimensions assume different properties\footnote{Borrowing the term from physics, where anisotropic is the opposite of isotropic or having a physical property which has the same value when measured in different directions.}?} Where would such differences come from? What would be the explanation for such anisotropy to appear in cases where the algorithms treat all dimensions in similar fashion? If structural bias can indeed possess such a property, any test for detecting it should necessarily take this fact into consideration, or otherwise be able to detect this kind of bias without assuming isotropy. 

This paper investigates the aforementioned questions based on results with such suspected anisotropy reported in \cite{Kononova2020CEC,Kononova2020PPSN}. The structure of the paper is as follows. 
In Section~\ref{sec:setup}, we describe the algorithms and structural bias detection mechanisms used in this paper. Section~\ref{sec:pot_anisotropy} introduces methods for detecting potential anisotropy, uses them to identify usecases, and tests whether their data actually shows anisotropy. Section~\ref{sec:test_impr} takes these usecases and identifies how they can be used to motivate new tests for structural bias. In Section~\ref{sec:concl}, we discuss these tests and their impact on future analysis of structural bias. 



\section{Experimental setup and data collection}\label{sec:setup}

\subsection{Structural bias detection}\label{sec:sb_detect}
Each algorithm under investigation is run $100$ times on function $f_0$ (Eq.~\eqref{eq:f0}) with $n=30$, which was first defined in~\cite{Kononova2015} to decouple the interaction between the objective function and the algorithm:
\begin{equation}\label{eq:f0}
    f_0:[0,1]^{n}\to[0,1], f_0(x)\sim\mathcal{U}(0,1).
\end{equation}  
Note that $f_0$ is truly stochastic - with values being randomly generated every time a position $x$ is evaluated - which means that its optimum is located uniformly in its domain $[0,1]^n$. Hence, an algorithm without structural bias is expected to produce uniformly distributed solutions in $[0,1]^n$.

By means of the visual~\cite{Kononova2015} and the statistical tests~\cite{Kononova2020CEC,Kononova2020PPSN}, we investigated, for some selected algorithms, whether the final points thereof follow a uniform distribution in $[0,1]^n$. The visual test was conducted by rendering the final points from multiple independent runs as a parallel coordinate plot, which we inspected visually against various clustering patterns within each dimension or across dimensions to determine the degree of structural bias. We further developed the statistical approach for automating such inspection process and for quantifying the structural bias. This approach involves testing each component/dimension of the final points against $\mathcal{U}(0,1)$ via the well-known Anderson-Darling (AD) test~\cite{Stephens74} and then aggregating all AD test statistics that produce significant decisions. Amongst various goodness-of-fit test procedures, e.g., the Kolmogorov-Smirnov (KS) and Cramer-Von Mises (CvM) tests~\cite{CsorgoJ96}, we chose the AD test due to its dominating statistical power over KS and CvM tests, which is obtained by simulating the alternative hypothesis with a mixture of beta distributions~\cite{Kononova2020CEC}. 

It is worth noting that this statistical approach is essentially a multiple comparison procedure, where we adopted the false discover rate (FDR) control to ensure a desired type-I error rate. This control inherently results in more conservative decisions (by generating less rejections) and thereby hampers the power of the overall statistical approach. Hence, in this paper, we counter this downside by taking a relatively large sample size. In Section~\ref{sec:ad_for_bouds}, we will discuss this in more detail.

\subsection{Algorithms and parameter settings}\label{sec:algsAndParam}
We consider a varied set of $41$ algorithms featuring population-based heuristics, Estimation of Distribution Algorithms (EDAs) and single-solution optimisation methods.

The population-based algorithms included to our experimentation are represented by $11$ widely used Differential Evolution (DE) variants \cite{Das2016}. These are employing the same values for the control parameters of their `compact' counterparts introduced in the next paragraph, but are executed with three different population sizes p$\in\{5,20,100\}$. By using common DE jargon, see \cite{Das2016,Caraffini2019,Kononova2020_outside} for details, the employed DE algorithms can be fully described with the nomenclature
\begin{itemize}
    \item \texttt{DE-best/1/bin} and \texttt{DE-best/1/exp};
    \item \texttt{DE/rand/2/bin} and \texttt{DE/rand/2/exp};
    \item \texttt{DE/best/2/bin} and \texttt{DE/best/2/bin};
    \item \texttt{DE/rand-to-best/2/bin} and \texttt{DE/rand-to-best/2/exp};
     \item \texttt{DE/current-to-best/1/bin} and \texttt{DE/current-to-best/1/ exp};
    \item \texttt{DE/current-to-rand/1}. 
\end{itemize}

Additionally, we consider $17$ so-called `compact' algorithms, i.e. EDAs mimicking the behaviour of established population-based algorithms through a simple probabilistic model where design variables are uncorrelated \cite{allcompacts}. A subgroup of these algorithms, namely  \texttt{cGA}, \texttt{cPSO}, \texttt{cBFO} and \texttt{cDElight} are taken with the same setup and parameters setting of \cite{Kononova2020PPSN}. Conversely, while keeping the same parameter setting of \cite{Kononova2020PPSN} (i.e. only two parameters are required regardless of the employed configuration), in this study we extend a number of `configurable' compact DE variants by equipping them with $13$ DE different combination of DE mutations and crossover operators, thus obtaining
\begin{itemize}
    \item \texttt{cDE/rand/1/bin} and \texttt{cDE/rand/1/exp};
\item \texttt{cDE/rand/2/bin} and \texttt{cDE/rand/2/exp};
\item \texttt{cDE/best/1/bin} and \texttt{cDE/best/1/exp};
\item \texttt{cDE/best/2/bin} and \texttt{cDE/best/2/exp};
\item \texttt{cDE/rand-to-best/2/bin} and \texttt{cDE/rand-to-best/2/exp};
\item \texttt{cDE/current-to-best/1/bin} and \texttt{cDE/current-to-best/ 1/exp};
\item \texttt{cDE/curr-to-rand/1}.
\end{itemize}


Finally, to round off the experimental setup with classic single-solution methods, we also consider:
\begin{itemize}
    \item An `iterated local search' referred to as the \texttt{RIS} algorithm.
    \item The two \texttt{Powell} and \texttt{Rosenbrock} deterministic methods. 
    \item The stochastic \texttt{Solis-Wets} and \texttt{SPSA}~\cite{Spall1987} algorithms, as well as its variant \texttt{SPSAv2} from \cite{Kononova2020CEC}.
    \item A `degenerative' single-solution particle swarm optimisation methods named \texttt{ISPO}.
    \item The `non-uniform' Simulated Annealing  \texttt{nuSA} algorithm as well as a standard Simulated Annealing \texttt{SA} with uniform distribution for the neighbouring operator and linear `cooling'. 
    \item The (1+1)--`Cholesky' covariance matrix Adaptation evolution strategy \texttt{(1+1)-CMAES}. 
    \item The two variants of the \texttt{(1+1)-ES} algorithm, see \cite{Kononova2020CEC} for details. 
    \item The popular `simplex' Nelder–Mead Algorithm (\texttt{NMA}). 
\end{itemize}
To run these $13$ single-solution methods, we maintained the parameters setting employed in \cite{Kononova2020CEC}.

Note that all $41$ employed algorithms are meant for general-purpose optimisation and are expected to explore the search space without showing preferential exploration directions along specific coordinate axes. 

\subsection{Strategies of Dealing with Infeasible Solutions}
Employing the most appropriate Strategy of Dealing with Infeasible Solutions (SDISs) is key, in particular when optimising several design variables as it is more probable to generate solutions outside the search domain \cite{Kononova2020_outside}. Hence, we execute all aforementioned algorithms with the $6$ SDISs reported below:  
\begin{itemize}
    \item Complete one-sided truncated normal strategy denoted as \texttt{COTN}.
    \item Dismiss strategy denoted here as \texttt{dis}.
    \item Mirror strategy denoted here as \texttt{mir}. 
    \item Saturation strategy denoted here as \texttt{sat}.
    \item Toroidal strategy denoted here as \texttt{tor}.
    \item Uniform strategy denoted here as \texttt{uni}.
\end{itemize}
This list forms an assorted set of SDISs, which has been fully described in \cite{Kononova2020PPSN,Kononova2020_outside}.
Note that these SDSIs are applicable for all algorithms under investigation excluding \texttt{cGA}, which only generates feasible solutions (and, thus, has the so-called inherent SDIS).
 
It is important to mention that that all the algorithms considered in this study and all SDIS treat all problem dimensions in an identical way.

\subsection{Experimental setup}\label{sect:setup}
\subsubsection{Data from optimisation runs} In order to study the structural bias of the algorithms under investigation, and its anisotropic behaviour, it is required to collect and store the position of the best found solution by each algorithm at the end of each performed run to then apply further processing and statistical analysis. This has been obtained by running the algorithms specified in Section \ref{sec:algsAndParam} with the parameters setting indicated in \cite{CaraffiniSOS_2020}. The source code for this entire experimental setup is made available at~\cite{code_algorithms_SB}, 
and a computational budget of $n\cdot 10000=300000$ fitness function evaluations.

Summarising, this means that this study considers:
\begin{itemize}
    \item $19$ amongst EDAs and single-solution algorithms employed with $6$ SDISs plus \texttt{cGA} (which does not require it);
    \item $11$ DE variants employed with $3$ population sizes and $6$ SDISs;
\end{itemize}
for a total of $19\cdot6+1+11\cdot3\cdot6=313$ algorithmic configurations to test.  

Note that only a subset of considered configurations is shown in figures throughout this paper -- configurations that do not exhibit any SB or suspected anisotropy based on the tests considered here are excluded from the figures.

\subsubsection{SB testing} The code used for processing the experimental data is available at~\cite{code_SB_anisotropy}. For the simulation of uniform random numbers, we use \verb|NumPy|~\cite{harris2020numpy}, and for the statistical tests we use their implementation in \verb|R|~\cite{R}, with the \verb|goftest|-package~\cite{goftest} for the AD-test.

\section{Potential anisotropy}\label{sec:pot_anisotropy}
\subsection{Identifying use cases}
\begin{figure}[!bt]
    \centering
    \includegraphics[width=0.47\textwidth, trim={20, 15, 0,  0}, clip]{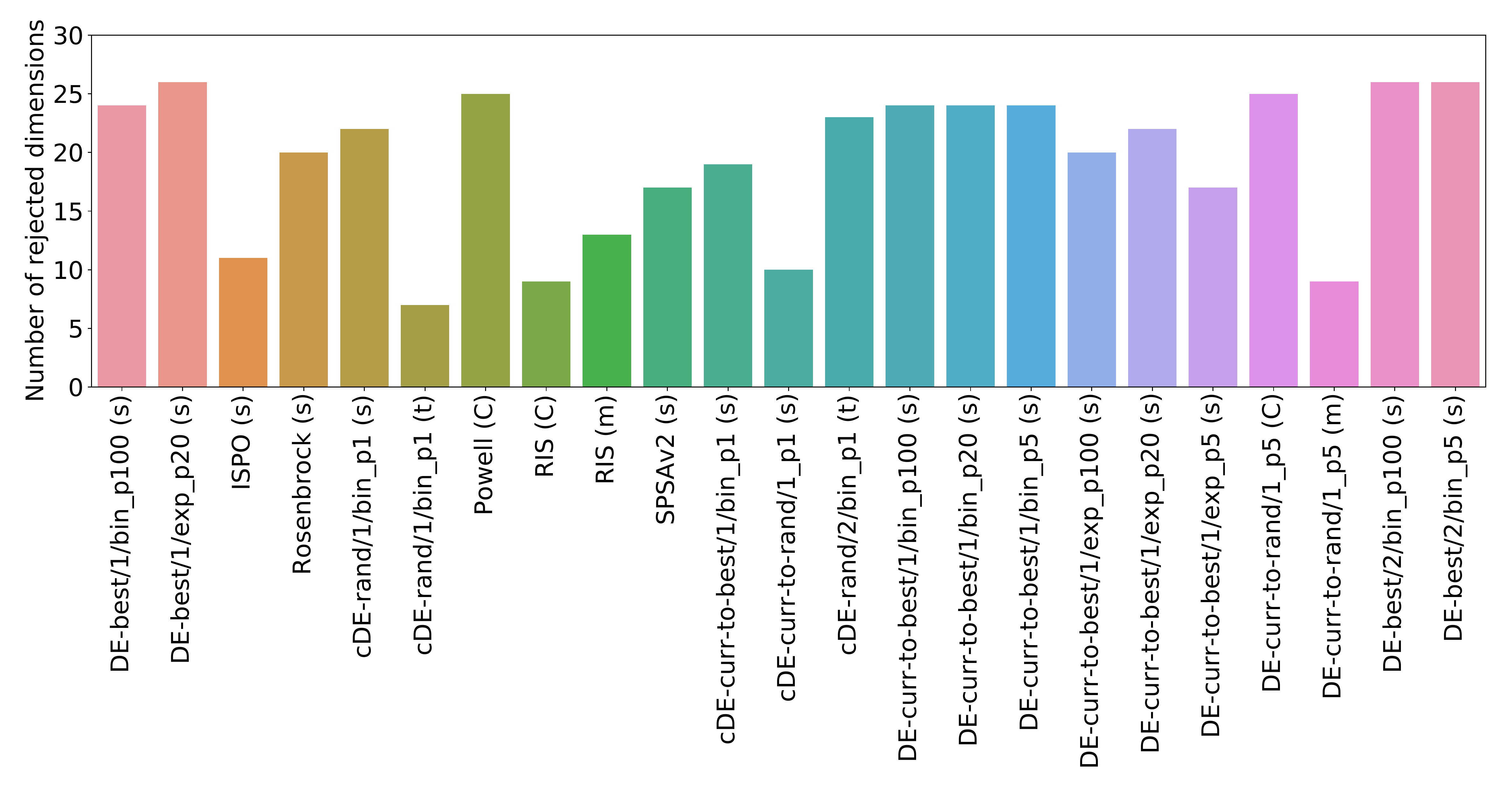}
    \caption{Number of rejected dimensions for different algorithm configurations based on the AD test described in Section~\ref{sec:sb_detect} ($\alpha =0.01$). 
    Only those with rejections $\in [4,26]$ are shown, both for readability and because the cases outside of these bounds could reasonably be explained as false positive / negative cases. }
    \label{fig:number_rejects_per_dim}
\end{figure}

We identify two main ways of detecting potential anisotropy in the final positions. The first is to consider the results of the previously described test for structural bias on a per-dimension basis. If an algorithm would be isotropic, we would expect this test to always lead to the same conclusion for each dimension. However, as we can see in Figure~\ref{fig:number_rejects_per_dim}, there are quite a few algorithms for which this does not seem to be the case. 

\begin{figure}
    \centering
    \includegraphics[width=0.47\textwidth, trim={20, 0, 0,  0}, clip]{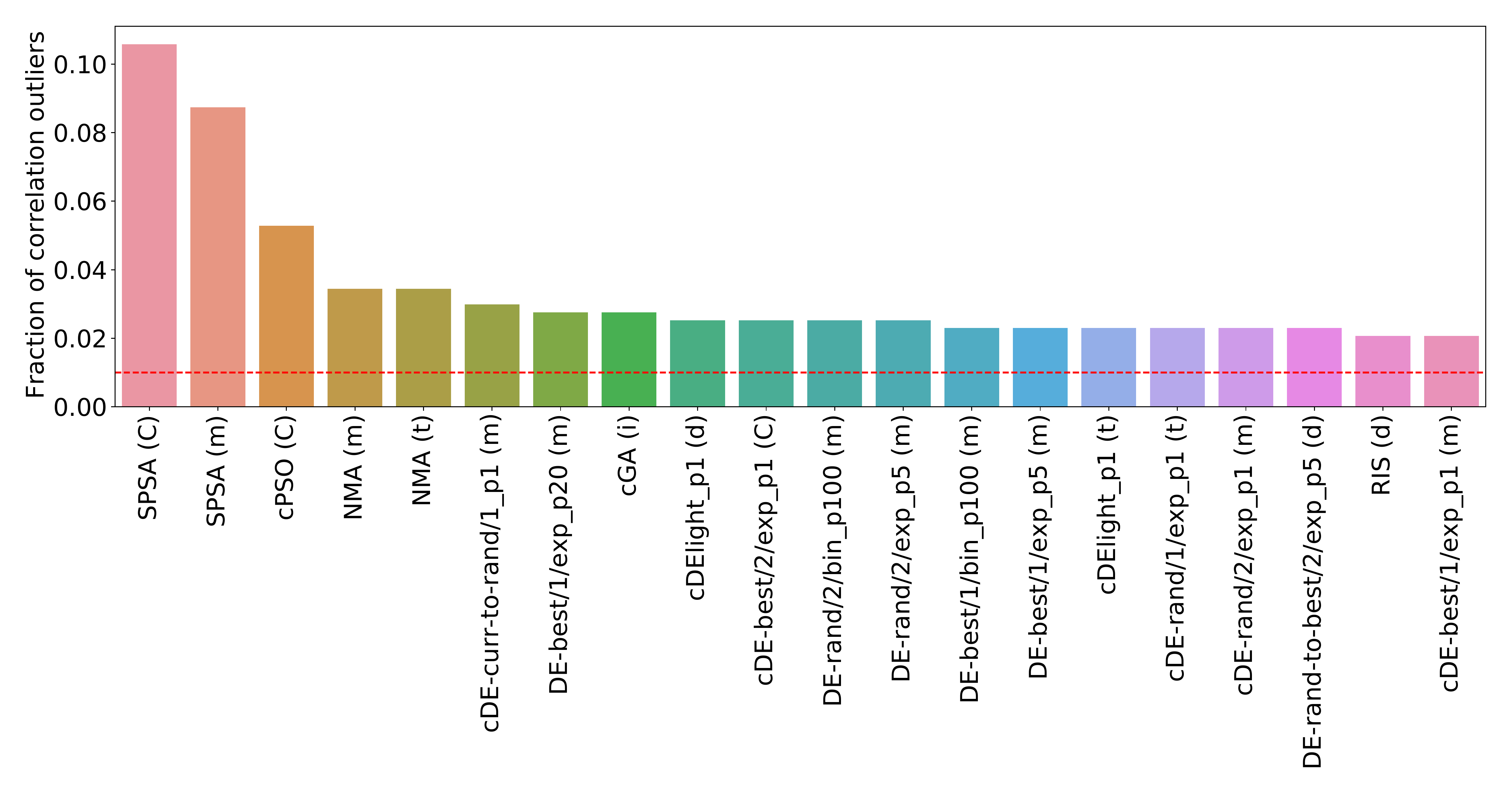}
    \caption{Fraction of pairwise Pearson correlation coefficients between dimensions which lie outside of the $0.99$ confidence bound of the independent uniform baseline. The subset of algorithm shown is based on the configurations for which the structural bias test did not reject any dimension, with the largest fraction of outliers. 
    The dashed line at $0.01$ represents the expected fraction of outliers resulting from uniform random distributions.}
    \label{fig:frac_outside_corrs}
\end{figure}

\begin{figure}
    \centering
    \includegraphics[width=0.45\textwidth, trim={20, 0, 0,  0}, clip]{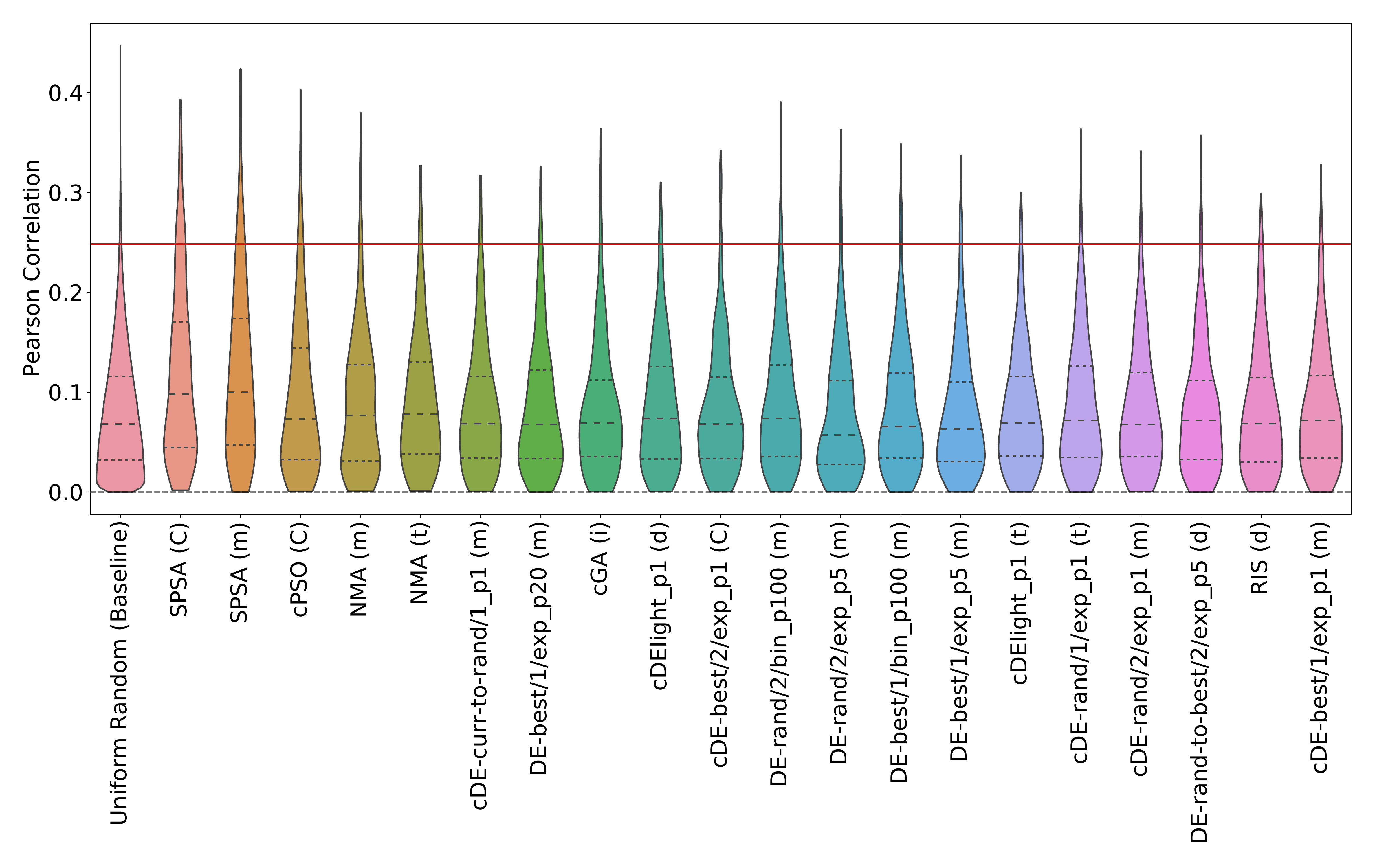}
    \caption{Pairwise absolute Pearson correlations between dimensions in the algorithms from Figure~\ref{fig:frac_outside_corrs}. The red lines represent the used boundary to calculate the fraction of outliers. The uniform baseline plot is calculated from 1000 repetitions of uniform random data (30 dimensions with 100 samples each, for every repetition). The red line corresponds to the 99th percentile calculated on the uniform samples.}
    \label{fig:correlation_plot}
\end{figure}

A second way in which potential anisotropy could be detected, is by looking at the correlations between the coordinates of the final positions. This can then be compared to the correlations we would expect from the case where all dimensions are indeed independent. This baseline is computed by simulating the correlation coefficients between independent uniform random variables in $[0,1]$. Since we are interested in the detection of outliers (pairs of dimensions with larger than expected correlation) 
relative to this baseline, we calculate the 99th percentile of absolute Pearson correlation between the coordinates of our random samples. Based on $10\,000$ simulations of uniform distribution, this value is $0.2484$.
We can then repeat this procedure for all considered configurations, resulting in $\frac{30\cdot29}{2}$ correlation coefficients. Then, we can say that the correlation is too large if the fraction of those above our threshold is larger than $0.01$ (selected to match the $\alpha$ value used throughout this paper). 
In Figure~\ref{fig:frac_outside_corrs}, we show this fraction for several configurations. 
The actual distributions of Pearson coefficients for these configurations is shown in Figure~\ref{fig:correlation_plot}. From this plot, it can be seen that the density outside of the calculated boundaries is indeed larger than uniform, but only a few of the cases are easily visually identifiable as deviating from random.

\subsection{Investigating Anisotropy}
While both of the above-mentioned methods show that there exist algorithm configurations which could be anisotropic in their behaviour on $f_0$, there are \textit{other factors which might explain these outliers as well}. To figure out if anisotropy is actually present, we need to look at the individual configurations in more detail. 

\begin{figure}
    \centering
    \includegraphics[width=0.48\textwidth, trim={100, 35, 125,  40},clip]{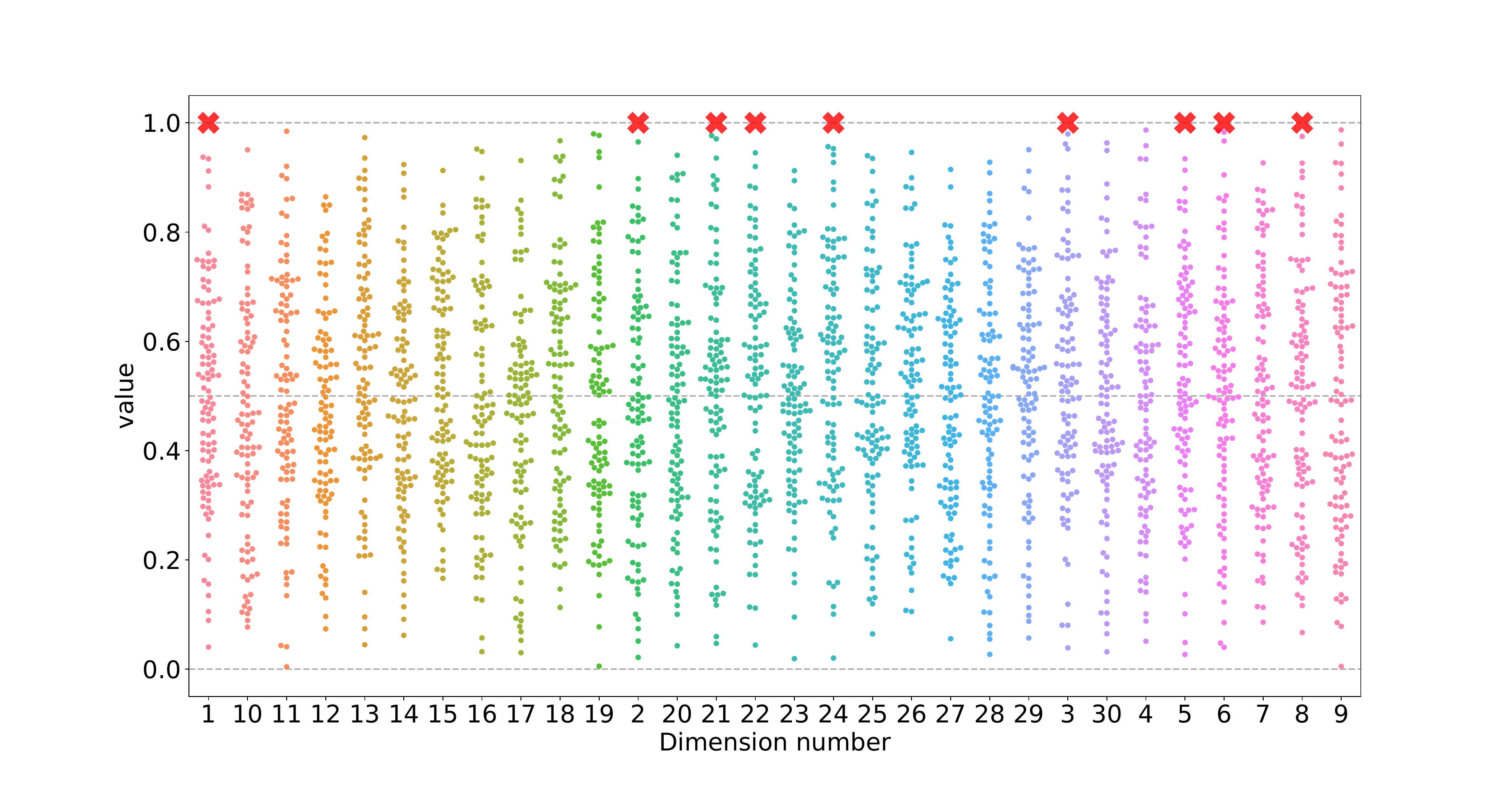}
    \caption{Swarmplot of the final positions found by algorithm configuration \texttt{DE-curr-to-rand/1-p5-mir}. The red crosses at the top indicate dimensions for which the original AD-test detects structural bias. }
    \label{fig:swarmplot_c622}
\end{figure}

First, we consider one of the configurations where the original test shows structural bias in some, but not all dimensions: \texttt{DE/curr-to-rand/1-p5-mir}. For this configuration, we plot the final positions in a parallel swarmplot\footnote{\url{https://seaborn.pydata.org/generated/seaborn.swarmplot.html}} in Figure~\ref{fig:swarmplot_c622}. This visualisation is similar to a scatterplot, but the points are adjusted in the categorical axis so they don't overlap. While this makes reading the exact y-values slightly harder, it provides a much clearer view of the distribution of points than a regular scatterplot. Visually, this plot shows a clear structural bias towards the centre, present in all dimensions. However, the AD-test only identifies issues in a small subset of them. This highlights an \textit{important shortcoming of the AD-test}: given our sample size of $100$, this test seems not to be able to accurately identify nonuniformity if the \textit{de facto} domain in a dimension is smaller than the domain of $f_0$.
We will study this deficiency in more detail in Section~\ref{sec:ad_for_bouds}.

\begin{figure}
    \centering
    \includegraphics[width=0.47\textwidth, trim={100, 35, 125,  40}, clip]{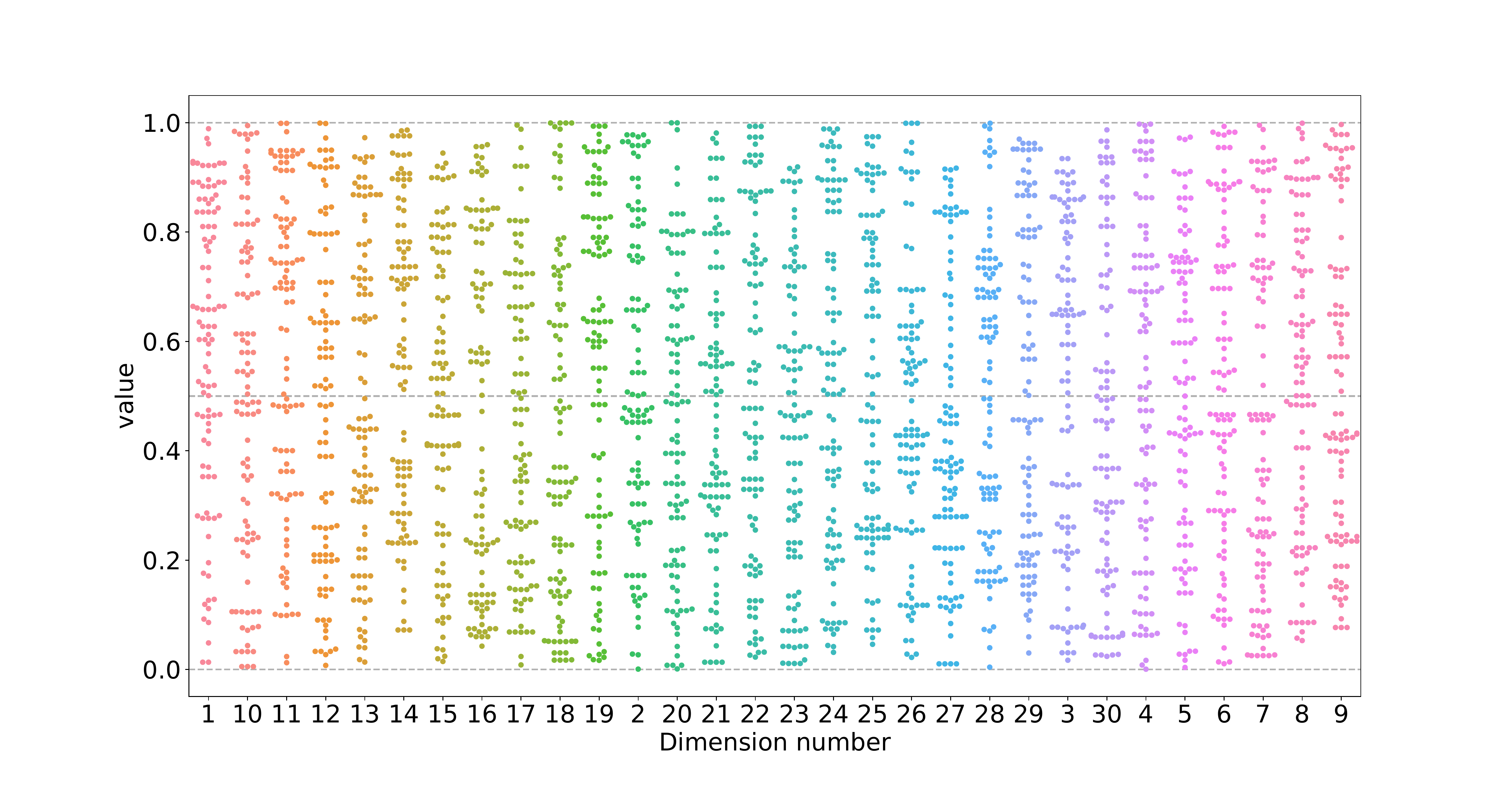}
    \caption{Swarmplot of the final positions found by algorithm configuration: \texttt{SPSA-COTN}. Note that none of the dimensions are rejected by the original SB detection procedure.}
    \label{fig:swarmplot_c466}
\end{figure}

Next, we look into one of the cases where we see a large number of seemingly correlated dimensions: \texttt{SPSA-COTN}. We show the swarmplot for this configuration in Figure~\ref{fig:swarmplot_c466}. To analyse the presence of anisotropy in this case, we use a 2-sample Kolmogorov-Smirnov (KS) test~\cite{massey1951kolmogorov} for each pair of dimensions, to check if there is any difference in their distribution. These p-values are then adjusted using the Benjamini-Yekutieli (BY) method~\cite{BenjaminiY01} to control the false discovery rate. When using $\alpha=0.01$, this procedure does not find any pair of dimensions for which we can say that they do not follow the same distribution. Thus, if anisotropy were present here, it can only be caused by correlations between dimensions. However, the test we used to check correlation makes the assumption that the data is uniformly randomly distributed in $[0,1]$. While the AD test did not reject this hypothesis, visual inspection of the swarm-plot raises doubts about the validity of this assumption. We clearly see heavy clustering, combined with relatively large empty gaps in all dimensions. Because of this, we should confirm the validity of the original testing procedure for finding structural bias. 


Better still, we could relax the assumption that the distribution of final points is uniform on $[0,1]$ through a permutation test, where for each dimension pair we repeatedly shuffle the data within each dimension and then compute the absolute Pearson correlation coefficients between them, resulting in a sample from the null hypothesis for this dimension pair (that they are not correlated). From this sample, we could approximate the critical value at the $\alpha$ level of significance, namely the $(1-\alpha)$th sample percentile. Now, consider the following collection of random indicators, $\mathds{1}(\rho_{ij} > c_{ij}), i = [1,..,n-1], j=[i+1,..,n]$, where $\rho_{ij}, c_{ij}$ are the observed correlation coefficient and the critical value for dimension pair $(i,j)$, respectively. Under an overarching hypothesis that there is no pairs of correlated dimensions, the sum of those indicator variables admits the following expectation: $\E\left\{\sum_{i,j}\mathds{1}(\rho_{ij} > c_{ij})\right\} = \alpha n(n-1)/2$, which could serve as a reference value/threshold for rejecting the overarching hypothesis, thereby indicating there is at least one pair of correlated dimensions (also the presence of anisotropy).

Even better, consider $\sum_{i,j}\mathds{1}(\rho_{ij} > c_{ij})$ (total counts of rejections) as a test statistic whose distribution under the overarching hypothesis can be approximated via bootstrapping, which draws a sample point by substituting $\rho_{ij}$ with a correlation coefficient calculated from a random permutation. 
From this bootstrapped sample, one can easily obtain a critical value at some level of significance, which can be used to compare to the observed counts of rejections for deciding whether to reject the overarching hypothesis or not (we could say that we leave this to the future work..).

\begin{figure}
    \centering
    \includegraphics[width=0.47\textwidth, trim={125, 55, 220,  40}, clip]{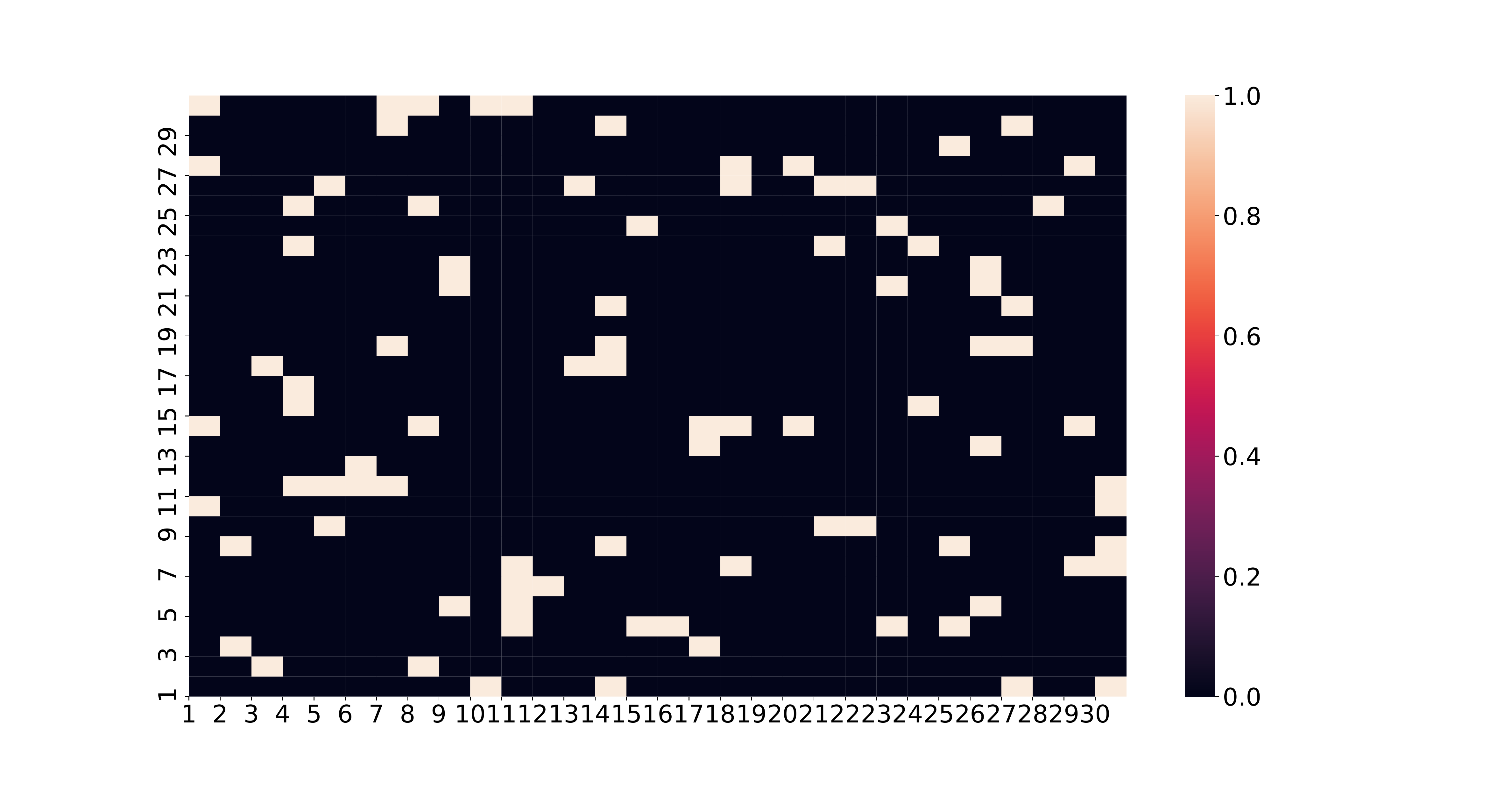}
    \caption{Heatmap for \texttt{SPSA-COTN} algorithm, showing which dimension pairs have Pearson correlation coefficient larger than expected based on the 99th 
    quantile of correlations between random permutations of the samples. Note that this matrix is symmetric and diagonal is ignored both in the plot and in the fraction-calculations.}
    \label{fig:relative_corr_matrix_c466}
\end{figure}

If we apply this procedure to the \texttt{SPSA-COTN} configuration from Figure~\ref{fig:swarmplot_c466}, we find that this test statistic gives us a fraction of $0.10$ dimension pairs which lie outside the critical range (with $\alpha=0.01$). The dimension pairs which are found to be potentially correlated by this approach are visualised in Figure~\ref{fig:relative_corr_matrix_c466}. Because of this, we can reasonably claim that this configuration is not fully isotropic. However, since our tests for distribution did not find any difference, this leaves \textit{open} the question of whether there is any structural bias present, which was just not found by the AD test procedure. 



\section{Improving the structural bias detection test}\label{sec:test_impr}
Based on the configurations investigated in Section~\ref{sec:pot_anisotropy}, we identified at least two scenarios in which the current structural bias detection technique does not match our visual inspection of the data (a higher density of points in the centre of the search space and a large number of clustered point in each dimension). 
In order to remedy this, we need to introduce new techniques which are more suited to tackle the particular types of deviations from uniformity that we have found. 

\subsection{Spacing-based test}\label{sec:spacing}
For the first type of deviation, where we observe a large amount of clustering, combined with large gaps which occur in all dimensions, we look at the distribution of the distances between consecutive points. We decide to focus only on this version of the spacing distribution, as opposed to more general $m$-spacings~\cite{pyke1965spacings}.
To decide whether or not to reject the samples in a particular dimension, we perform a 2-sample KS test against a set of spacings from $1000$ samplings of the uniform distribution. Note that this does include the distance to both boundaries of the dimension. 

\begin{figure}
    \centering
    \includegraphics[width=0.47\textwidth, trim={135, 15, 100,  10}, clip]{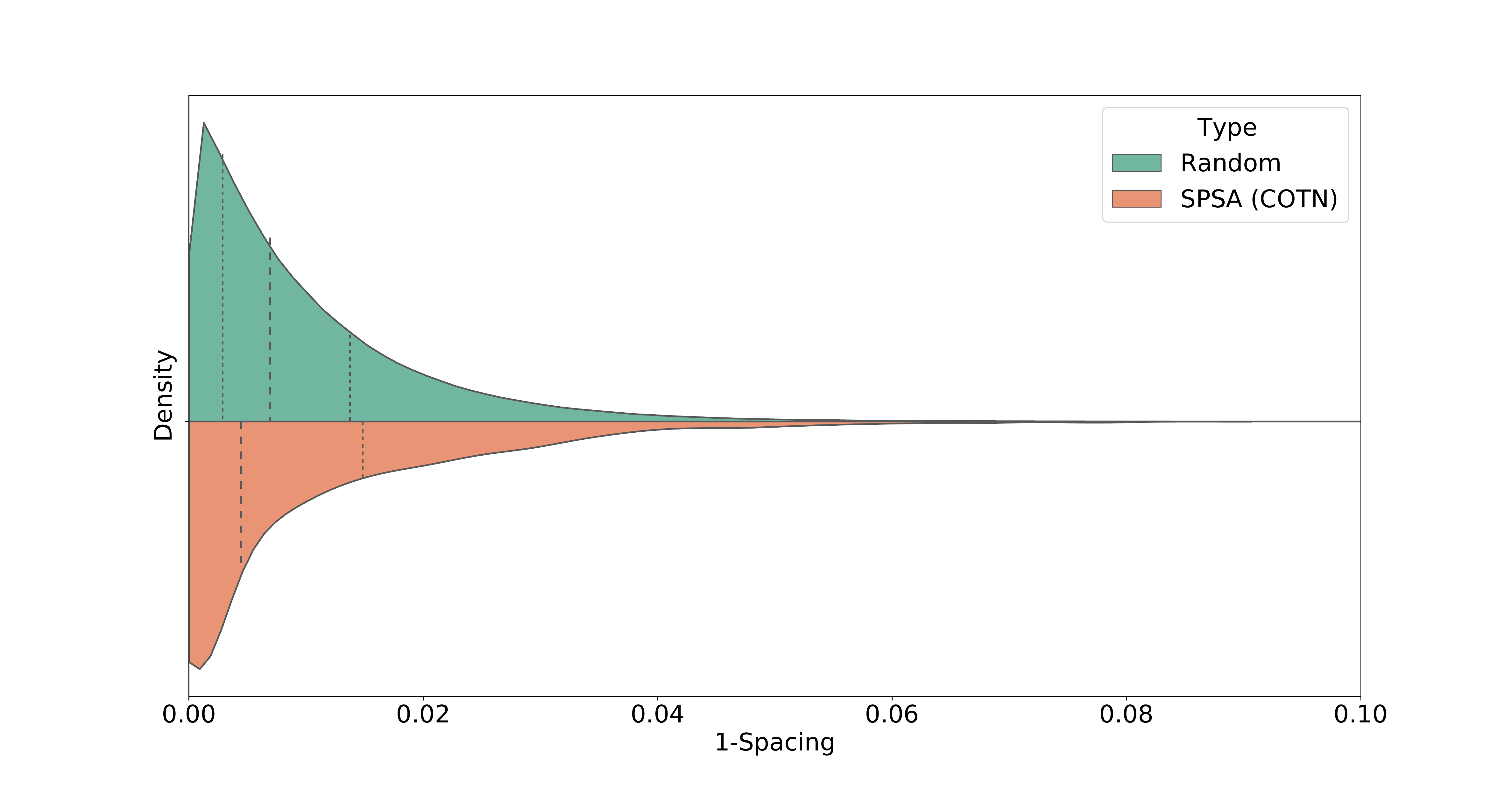}
    \caption{
    1-spacing statistic aggregated over all 30 dimensions of the positions found by \texttt{SPSA-COTN}, against the baseline distribution based on $10000$ repetitions of $100$ uniform random samples in $[0,1]$. }
    \label{fig:violin_spacing_c466}
\end{figure}

This spacing-test can be done on each dimension individually, or on the aggregation among all dimensions. To show the effectiveness of the test, we plot this aggregation for the configuration which was discussed in Figure~\ref{fig:swarmplot_c466}, \texttt{SPSA-COTN}. The aggregation is compared to the baselines described above, and shown in Figure~\ref{fig:violin_spacing_c466}. This figure shows that the differences in distribution are large, indicating that the positions found by this SPSA variant show a clear deviation from uniformity, which could not be detected using the original AD-test.

\subsection{Boundary-Related Structural Bias}\label{sec:ad_for_bouds}
To create a test which is more reliable in cases where algorithms show bias towards the centre of the space, we propose to perform a simple transformation on the input samples, collapsing them across the centre into the domain $[0, \frac{1}{2}]$, and performing an AD-test on these transformed samples. Note that this transformation inherently decreases the effectiveness of the test on cases where an asymmetry with regard to the centre is present, so this test should not replace the standard AD-test, only complement it. 

We can illustrate the effectiveness of such transformation by considering a simplified version of a structurally biased dimension: data which is sampled from a uniform distribution in $[0.05, 0.95]$. If we simulate both SB detection procedures on $1000$ times on $100$ uniform samples following this modified domain, we find that the original test only rejects in $2.2\%$ of cases, as opposed to $24.9\%$ for the transformed case ($\alpha=0.01$). While this is still \textit{nowhere near} optimal ($100\%$), it is a large improvement. This shows when we run the test on the data from Figure~\ref{fig:swarmplot_c622}: the test on the transformed samples rejects for all dimensions. 
\begin{figure}
    \centering
    \includegraphics[width=0.47\textwidth, trim={100, 35, 125,  40}, clip]{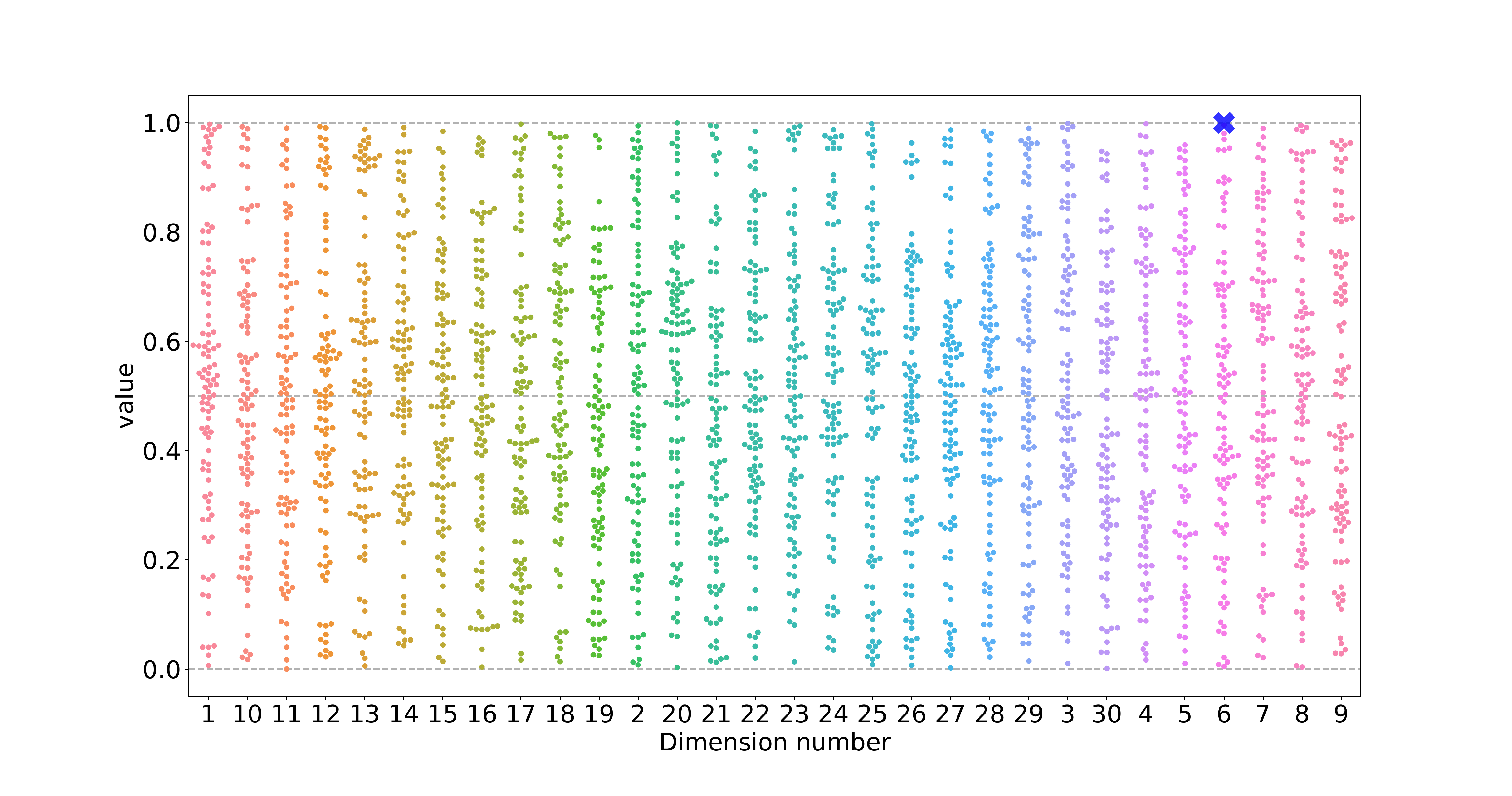}
    \caption{Swarmplot of the final positions found by algorithm configuration: \texttt{cPSO-COTN}. The blue cross(es) at the top indicate dimensions for which the transformed AD-test detects structural bias.}
    \label{fig:swarmplot_c486}
\end{figure}

\subsection{Aggregated Tests}
\begin{figure}
    \centering
    \includegraphics[width=0.47\textwidth, trim={125, 35, 100,  30}, clip]{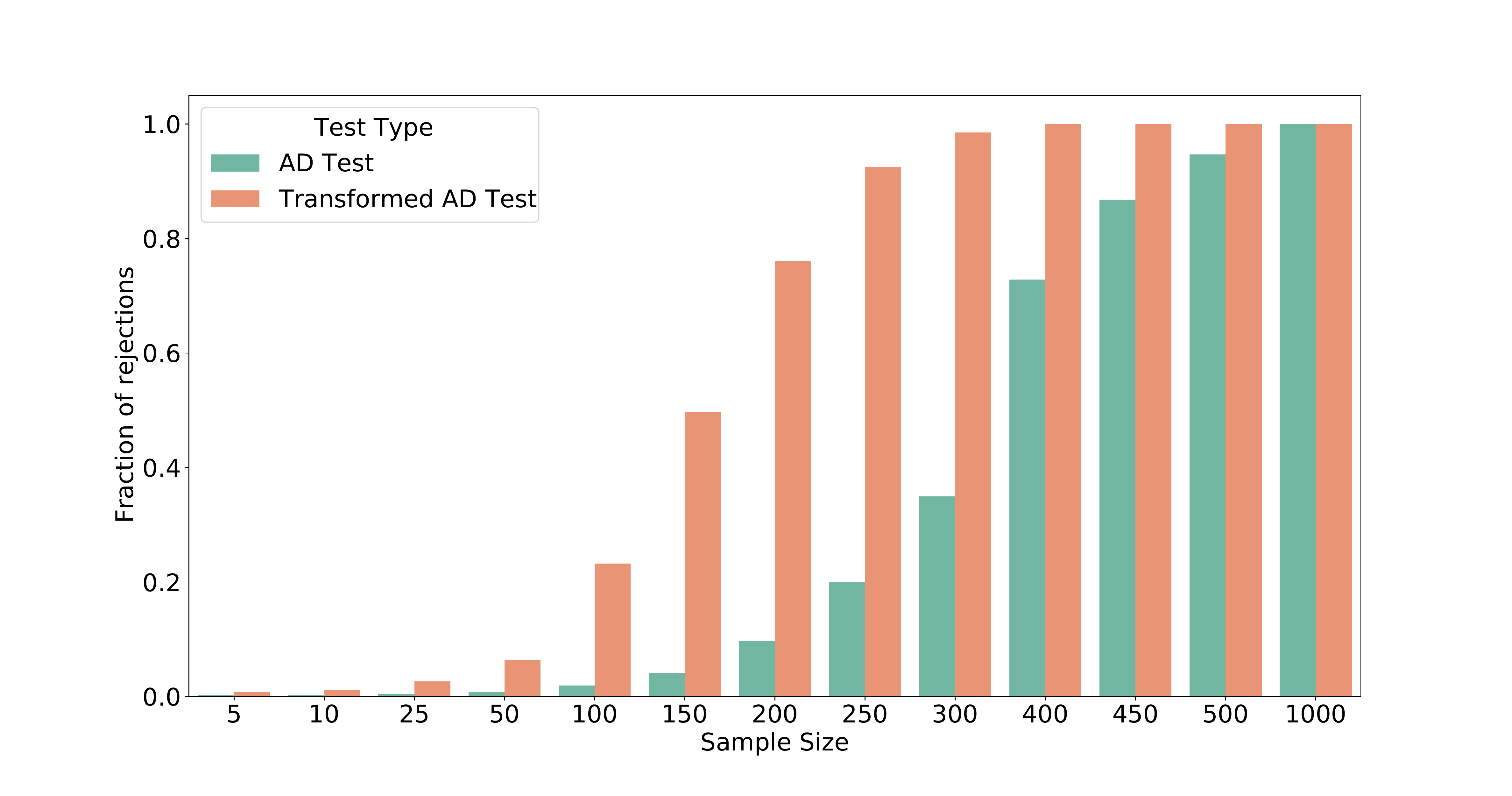}
    \caption{Overview of the fraction of samples from the uniform distribution on $[0.05,0.95]$ is rejected by the two versions of the AD test based on the size of the sample ($\alpha=0.01$). Fractions calculated based on $10\,000$ simulations. Transformed AD test refers to the AD test on the transformed samples as discussed in Section~\ref{sec:ad_for_bouds}}
    \label{fig:samplesize}
\end{figure}

While performing these tests on each dimension individually is recommended in general case where an algorithm can treat dimensions differently, it does lead to a scenario where each dimensions sample size might be too small to detect any structural bias, even though it can be found for the data as a whole. We can show this by considering another case from Figure~\ref{fig:frac_outside_corrs}: \texttt{cPSO-COTN}. For this configuration, the original test found no deficiencies. However, with visual inspection, as shown in Figure~\ref{fig:swarmplot_c486}, we would suspect SB to be present. Indeed, if we change the testing procedure to perform the AD-test on the full collection of samples (since the codomain of $f_0$ is the same in all dimensions, this does not require any transformation), so aggregated over all 30 dimensions, it would reject the null-hypothesis of uniformity ($\alpha=0.01$). This highlights the fact that while the additional tests suggested in this section can be useful to find more types of SB, they still struggle to detect more subtle cases, since the sample size of 100 is not large enough for these testing procedures. 
We further illustrate this point by simulating different samples in the uniform distribution on the domain $[0.05,0.95]$ and checking what fraction is rejected by these tests. We repeat these simulations $10000$ times for each sample size, and show the results in Figure~\ref{fig:samplesize}. From this figure, we can see that even with the AD test on the transformed samples, we need more than $300$ samples per dimension to reject samples from this distribution with a reduced range (without taking into account any subsequent p-value correction). However, visually, these cases are easy enough to identify, since the inaccessible part of the domain is 10\% of the complete space. And indeed, aggregating samples over dimensions provides enough samples for the original AD test to reject the null hypothesis. 
It is worth noting that the transformation of samples in this way is specifically designed to be more effective in cases where the deficiency is shaped like this, and not a generic technique. As such, the sample-sizes shown here are purely illustrative, and \textit{will} depend on the particular set of test and types of SB to detect. 

\begin{table*}[]\small
\caption{Overview of the different SB detection methods discussed in this paper and the types of SB they can identify.}\label{tab:sb_methods}
\begin{tabular}{@{}lll@{}}
\toprule
Method & Description & Type of SB detected \\ \midrule
Default AD test & Per dimension AD test of uniformity in $[0,1]$ & As described in~\cite{Kononova2020PPSN} \\
AD test on Transformed Samples & \begin{tabular}[c]{@{}l@{}}Per dimension AD test of uniformity of transformed \\ samples ($|x-\frac{1}{2}|$) in $[0,\frac{1}{2}]$\end{tabular} & \begin{tabular}[c]{@{}l@{}}More effective in finding bias towards\\ or away from centre of search space\end{tabular} \\
1-Spacing test & \begin{tabular}[c]{@{}l@{}}Per dimension test on distances between (sorted) points,\\  including the boundary. Tested using 2-sample KS test\\  against simulated random samples.\end{tabular} & Clustering of points in each dimension \\
AD test on aggregated samples & AD test of uniformity in $[0,1]$ on all dimensions at once & Overall indication on presence of SB \\ \bottomrule
\end{tabular}
\end{table*}

\begin{figure*}
    \centering
    \includegraphics[width=\textwidth]{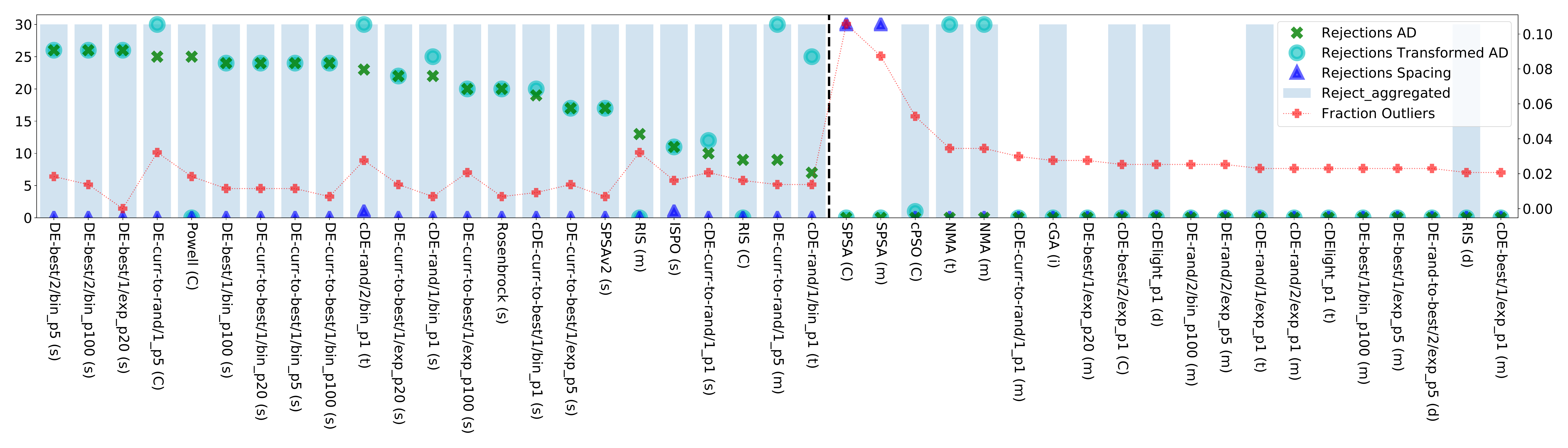}
    \caption{This figure shows the results of the proposed statistical SB detection methods on the configurations where the original test doesn't give the same result for each dimension, shown in Figures~\ref{fig:number_rejects_per_dim} and left of vertical dashed line here, and configurations where the fraction of outliers in the pairwise correlation between dimensions is larger than expected, shown in Figure~\ref{fig:frac_outside_corrs} and right of vertical dashed line here. The rejection numbers are shown on the left axis, while the fraction of correlations outside the expected range is read on the right axis. The aggregated rejections is binary, so is the bar is present the aggregated AD test reject the null hypothesis ($\alpha = 0.01$). The configurations are sorted based on number of rejections by the original AD test, and by fraction of outliers in case of a tie.}
    \label{fig:overview_all_methods}
\end{figure*}

In order to get a more complete overview of the effect of the different proposed techniques for detecting SB, we summarise the testing procedures in Table~\ref{tab:sb_methods}. Additionally, we have run them on all algorithm configurations shown so far. Based on this, we have created an overview in Figure~\ref{fig:overview_all_methods}. From this visualisation, we can see that the aggregated AD test in particular identifies a lot more configurations as structurally biased than the original per-dimension AD test. This highlights the dependence of the AD test on a relatively large sample size to detect the relatively minor deviations from the standard uniform distribution that can easily be seen when visualising the data in swarm or scatter-plots. We also see that the 1-spacing test detects only a few cases of SB, but the cases it detects are not found accurately by any of the other methods, indicating that is should still be considered an important part of a future portfolio of SB-tests.
Additionally, we see that for some of the configurations with only a slightly larger than expected fraction of outliers, none of the described tests find any reason to reject the null-hypothesis. This can either indicate a lack of bias, in which case the sensitivity of the method used to define outliers might be too large (considering there is no form of multiple-test correction in this procedure) and there is no bias present. Alternatively, these configurations are only mildly biased, in which case our detection methods are still not sensitive enough. Based on the techniques discussed in this paper, we can't definitively give a conclusion one way or the other. Further study is required to be able to effectively differentiate and quantify different types of bias, especially in cases where this bias is mild. 





\section{Discussion and future work}\label{sec:concl}
Based on the results presented in Section~\ref{sec:pot_anisotropy}, we can say with confidence that very few of the algorithm configurations considered here show signs of structural bias. With the exception of the SPSA algorithm, all identified cases of potential anisotropy can be explained by the shortcomings of individual AD-test. Indeed, even for this SPSA algorithm, we have identified that structural bias is present which can be detected by tests which are not impacted by anisotropy, leading us to conclude that anisotropy has no negative impact on the ability to detect the presence of SB. 

Based on the use cases we identified, we have introduced several additional testing procedures for detecting structural bias. While this proves to be a challenging problem, these new tests are able to detect different types of SB which were missed by the original approach. Indeed, we have been able to show that the cases where the results of this original SB detection could be interpreted as showing anisotropy in the algorithm configurations are instead the result of a lack of sensitivity to several types of SB. 
These results clearly indicate that different types of SB require different testing procedures. As such, a thorough investigation into these different types of deficiencies is needed to design effective tests. As we have shown for the case where part of the domain is not reached by the algorithm, different tests will show different behaviour for varying sample sizes. In order to effectively test for SB, the relationship between sample size, test procedure and type of detected SB has to be made explicit. 

With a more detailed understanding of the kinds of SB and methods to detect them, one should be able to construct a portfolio of testing procedures, which can be used to check any algorithm for signs of SB. This portfolio should then take into account limitations on sample size, and provide the most effective testing setup for the users constraints, leading to a measure characterising the presence of SB, and if so, which type and to what extent. This would be highly useful for algorithmic design (while introducing the algorithmic operators), as detecting and characterising SB early on in the design process can give useful insights into the impact of individual algorithmic operators when they are introduced, and on their collective behaviour. This information can lead to a better understanding of the algorithm and its operators, which is an important step on the path towards removing structural deficiencies from iterative optimisation heuristics.

\balance

\bibliographystyle{ACM-Reference-Format}
\bibliography{paper}

\appendix




\end{document}